\documentclass[onecolumn,amsmath,amssymb,aps,prd,showpacs,showkeys,nofootinbib,floatfix]{revtex4}

\usepackage[dvips]{graphicx}
\usepackage[dvips]{color}

\newcommand{\be}{\begin{equation}}
\newcommand{\ee}{\end{equation}}

\def\fun#1#2{\lower3.6pt\vbox{\baselineskip0pt\lineskip.9pt
\ialign{$\mathsurround=0pt#1\hfil ##\hfil$\crcr#2\crcr\sim\crcr}}}

\newcommand{{\SD}}{\rm SD}

\newcommand{\vexi}{\mbox{\boldmath${\rm \xi}$}}
\newcommand{\veta}{\mbox{\boldmath${\rm \eta}$}}

\newcommand{{\Mc}}{\mathcal{M}}

\newcommand{\lan}{\langle}
\newcommand{\ran}{\rangle}

\begin{document}

\author{Yu.A.Simonov}
\email{simonov@itep.ru}

\affiliation{ITEP, Moscow, Russia}

\title{Theory of time-like baryon form factors near thresholds}

\date{}

\begin{abstract}
A new mechanism of baryon-antibaryon production via nonperturbative double pair
creation in   intermediate mesons is proposed and the theory contains no
fitting parameters. It is shown, that near-threshold resonances are responsible
for enhancements in electroproduction  cross sections: $\psi(4S)$ for
$\Lambda^+_c \Lambda^-_c, \Upsilon (6S)$ for $\Lambda_b \bar \Lambda_b$.  An
admixture of intermediate $D$- wave resonances produces angular dependence in
differential cross section and
  can explain  unusual
behavior of the ratio $G_E/G_M$ for the proton.

\end{abstract}


\maketitle



\section{Introduction}

The topic of time-like baryon form factors and barion-antibaryon
$(\mathcal{B}\bar { \mathcal{B}})$ production is being actively studied
experimentally for the last 30 years, for reviews and  and references see
\cite{1,2}. As a result an   impressive amount of data  on   $p\bar p,
\Lambda\bar \Lambda, \Sigma\Lambda,  \Sigma\Sigma, \Lambda_c\Lambda_c$ is
obtained, signalling strong enhancements  in all cases near the corresponding
thresholds. Moreover, in some cases the angular dependence of cross sections
was obtained \cite{3,4}, which allows to distinguish between $G_E$ and  $G_M$
and obtain its ratio, different from unity near threshold. These features are
well established and call for explanation. On theoretical side two approaches
are most popular. In the   first one exploits perturbative approach with model
assumptions about distribution amplitude  functions, appropriate for  higher
energies \cite{5}. A quark-diquark model   with more parameters was given in
\cite{6} and a more phenomenological approach  suggested in \cite {7}.

In the second type of approach the $ \mathcal{B} \bar{ \mathcal{B}}$  threshold
enhancements are treated as due to molecular or four-quark mechanisms, see
\cite{8} for a recent work in this direction.

In this letter we propose a new mechanism of  $ \mathcal{B} \bar{ \mathcal{B}}$
production and structure formation in time-like form factors, which is of
universal  character and is free from  fitting parameters. This mechanism is
based on the detailed description of the  string breaking with creation of two
light quark pairs, e.g. $(u\bar u), (d\bar d)$. The  relativistic theory of
string breaking with one pair $(q\bar q)$ creation was  given in \cite{9} and
depends on the only parameter -- string tension $\sigma$.

\begin{figure}
\caption{Elecrtoproduction of $\mathcal{B}\bar{\mathcal{B}} $ by double string
breaking mechanism}
\includegraphics[width=70mm,keepaspectratio=true]{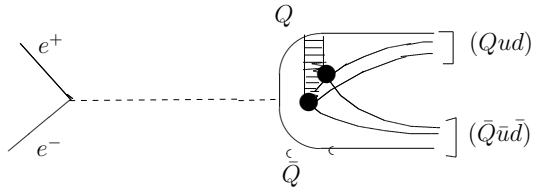}
\end{figure}

In case of double pair creation the initial state is again a meson $(Q\bar Q)$
as shown in Fig.1 and the string breaking vertex contains in addition to
$\sigma$ the fundamental parameter -- the vacuum correlation length $\lambda$,
which was measured both on the lattice and analytically, $\lambda\approx (0.1
\div 0.2)$ fm \cite{10}. The  $ \mathcal{B} \bar{ \mathcal{B}}$
electroproduction cross section was calculated in \cite{11} in the case, when
$(Q\bar Q)$ is in the $(n^3S_1$) - state and its wave function $\psi_n(r)$  is
represented by SHO functions  with SHO parameter $\beta_n$, whereas for baryons
the  lowest hyperspherical mode of gaussian form was used, $\Psi_{\mathcal{B}}
\sim \exp \left( -\frac{\vexi^2 + \veta^2}{R^2_0}\right), ~~ R^2_0 = \frac16
\lan r^2_B\ran, (\vexi, \veta$ are normalized Jacoby coordinates, see \cite{12}
for details). The resulting electroproduction cross section can be written as a
double  sum over $Q\bar Q$ and its excited states, \be \sigma (e^+e^- \to
\mathcal{B} \bar{ \mathcal{B} }) = \frac{12\pi \alpha^2 p\lambda^4}{E^3} \left|
\sum_Q e_Q \sum_n \frac{\psi_n (0) \eta_{\mathcal{B} Q} J_{n   \mathcal{B}
\bar{ \mathcal{B}} }(p)}{E_n - E - i\frac{\Gamma_n}{2}}\right|^2\label{1}\ee
where $E$ is the total c.m. energy , $p$ -- baryon c.m. momentum and $J_{n
\mathcal{B} \bar{ \mathcal{B}}} (p)$ is the overlap integral of $\psi_n(r)$ and
$\mathcal{B}\bar{\mathcal{B}}$ wave functions, given in Eq. ({29}) of
\cite{11}. Here  $\eta_{\mathcal{B} Q}$ is a spin recoupling coefficient
$\eta_{\Lambda_cc}=1, ~~ \eta_{pu} =\frac43$.

In case, when only one $(n^3S_1)$ state is close to the threshold for a given
pair $(Q\bar Q)$ and  the only pair $(Q\bar Q)$ is dominant, as in the case of
$\Lambda_c \bar \Lambda_c$ or $\Lambda_b \bar \Lambda_b$ production, the
behavior of $\sigma$ can be written as \be \sigma (e^+e^- \to \mathcal{B} \bar{
\mathcal{B}} )= C_n \frac{p}{E^3} \frac{\exp (-p^2R^2_0 \bar c)}{(E-E_n)^2 +
\frac{\Gamma^2_n}{4}},\label{3}\ee where the constants $C_n, \bar c$ are
calculable through parameters of baryons and $(n^3S_1)$ state. The latter are
easily   calculated in the relativistic Hamiltonian, comprising the quarks with
current masses $m_q(m_u=m_d\approx 0; m_s\approx 0.17$ GeV, $m_c =1.42 $ GeV,
$m_b = 4.83$ GeV, the QCD string between $Q$ and $\bar Q$ with $\sigma=0.18 $
GeV$^2$ and $\alpha_s$ with asymptotic freedom at small and freezing at large
distances. For high excited states also the coupling to decay channels is
important, which is taken into account by the so-called flattening potential.
The calculated  parameters for charmonia \cite{13}, bottomonium \cite{14} and
light mesons  are given in the Table, together with PDG data \cite{14'}.

The   resulting behavior of cross section for  $e^+e^-\to \Lambda_c\bar
\Lambda_c$ is shown in Fig. 2, together with experimental data from \cite{15}.
Theoretical prediction used $4^3S_1$ charmonium state with parameters shown in
Table and $\lambda=0.2$ fm, $ \lan r^2_{\Lambda_c}\ran = (0.8$ fm$)^2$. A more
accurate fit to the data (shown in Fig. 2 by broken line) requires $\lambda$ to
be 15\% smaller.

\begin{figure}
\caption{{ The cross section $ \sigma (e^+e^-\to \Lambda^+_c \Lambda^-_c)$ in
$nb$,  with $\lambda=1$ GeV as function of of $M(\Lambda^+_c \Lambda_c^-) = E/
c^2$  (solid line) experimental points are from \cite{15}, dashed line is the
best normalization fit of Eq. (\ref{3}) with $\lambda=0.17$ fm.}}
\includegraphics[width=60mm,keepaspectratio=true]{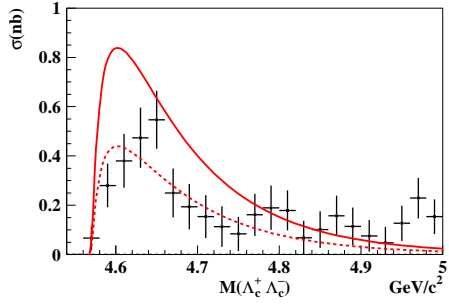}
\end{figure}

\begin{figure}
\caption{$\Lambda_b \bar \Lambda_b$  electroproduction cross section  in $nb$
near threshold as a function of total energy $E$.}
\includegraphics[width=60mm,keepaspectratio=true]{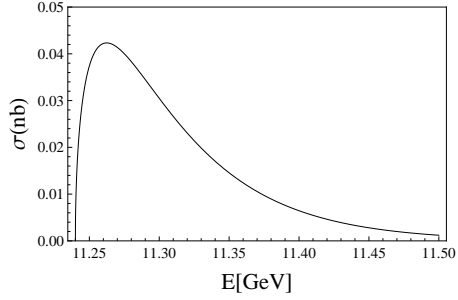}
\end{figure}

The case of  $\Lambda_b \bar \Lambda_b$ production is treated in the same way,
with  $6^3S_1$ $\Upsilon(11.02)$ as an intermediate state. Using parameters
from the Table, and the same $\lambda$ and $R_0$, one obtains the curve, shown
in Fig.3. One can see again the threshold enhancement, produced by the form of
Eq. (\ref{3}). A similar analysis of $ \Lambda \bar \Lambda$,  $p \bar p$
electroproduction should take into account angular distribution, which was
measured in \cite{16} and \cite{4} respectively. To this end one should
 include angular dependence due to $D$ wave admixture in a vector
$1^{--}$ resonance with the wave function \cite{17} \be \Psi_n \equiv
\frac{e_i}{\sqrt{8\pi}} v^+ (\sigma_i R_{nS} (r) \xi_{nS} + \mathcal{P}_{ik}
\sigma_k \xi_{nD}  R_{nD} (r) )v,\label{4}\ee where $ \mathcal{P}_{ik}
=\frac{3}{\sqrt{2}} (n_in_k -\frac13 \delta_{ik})$, and $|\xi_{nS}|^2 +
|\xi_{nD}|^2=1$. Insertion of (\ref{4})  into $J_{n \mathcal{B}
\bar{\mathcal{B}}} (p)$ yields a $D$-wave part of the latter, $J_{n \mathcal{B}
\bar{\mathcal{B}}} (p)\to J_{n \mathcal{B} \bar{\mathcal{B}}}^{(S)}+J_{n
\mathcal{B} \bar{\mathcal{B}}}^{(D)}$, and the resulting differential cross
section can be written as

\begin{table}
\caption{}\label{1}
\begin{center}
\begin{tabular}{|l|l|l|l|l|}\hline
 $Q\bar Q$& \multicolumn{1}{c|}{ $b\bar b$ } &\multicolumn{1}{c|}{$c\bar c$}&\multicolumn{2}{c|}{ $u\bar u\pm d\bar d$}\\
 \hline nS&6S&4S&3S&2D\\\hline
\hfill th&11.04& 4.426&1.9&1.99\\
  $ E_{n}$(GeV)&&&&\\
\hfill [15] exp&11.023&4.421&1.9&$\sim 2$\\\hline

 $\psi_n(0)$ (GeV$^{3/2}$)&0.38&0.13&0.13&0.0275\\\hline

 $\beta_n$(GeV)&0.55&0.43&$\sim 0.3$&$\sim 0.3$\\ \hline
 $\Gamma_{n}$ (MeV)[15]&$79\pm 16$& $62\pm 20$&$130\div160$&$\sim
 150$\\\hline
 $\bar C_{n}( nb)$ &14.1&7.7&&\\\hline
 $\bar cR^2_0$ (GeV$^{-2}$)&2.5&2.5&&\\\hline

\end{tabular}
\end{center}
\end{table}

\be \begin{gathered}\frac{d\sigma(e^+e^-\to \mathcal{B}
\bar{\mathcal{B}})}{d\Omega} = \alpha^2\rho^2\frac{p}{E^3} e^{-p^2R^2_0 \bar
c}\left\{ |\Xi_S|^2 + \frac{1}{\sqrt{2}} Re
(\Xi_S\Xi_D^*)+\right.\\
\left.+\frac54 |\Xi_D|^2-\cos^2\theta \left(\frac34 |\Xi_D|^2 +
\frac{3}{\sqrt{2}}Re (\Xi_S\Xi_D^*)\right)\right\},\end{gathered}\label{5}\ee
with
\be \Xi_A = \sum_n \xi_{nA} \psi_{nA} (0) \frac{Q_{nA} (p)}{E -E_n+ \frac{i
\Gamma_n}{2}},~~~~ A=S, D\label{6}\ee and $Q_{nD}(p)$ depends on $R_0,\beta_n$
and is a polynomial in $p^2$

 \be \rho^2 =96 \pi^{3/2} e^2_Q~ \eta_{BQ}^2 \lambda^4, ~~ \psi_{nS}(0) = \frac{1}{\sqrt{4\pi}}
R_{nS} (0), ~~ \psi_{nD} (0) = \frac{5 R^{''}_{nD}(0)}{4\sqrt{8\pi}
\omega^2_Q}.\label{7}\ee

Here $\omega_Q=\lan \sqrt{p^2+m^2_Q}\ran_n$. Note, that at small $p$ the
polynomial $Q_{nD} (p) = O\left(\frac{p^2}{\beta^2_0}\right)$, hence $\Xi_D
(p\to 0) \sim O(p^2)$. Comparison to the standard definition of (modified)
Sachs form factors $G^P_M (E)$ and $G^P_E(E)$, immediately yields \be
\left|\frac{2M}{E} G^P_E\right|^2 = \rho^2 e^{-p^2R^2_0\bar c} \left\{
|\Xi_S|^2 + 2\sqrt{2} Re (\Xi_S \Xi_D^*) + 2 |\Xi_D|^2\right\}\label{9}\ee \be
|G^P_M|^2 =\rho^2 e^{-p^2R^2_0\bar c} \left\{ |\Xi_S|^2 - \sqrt{2} Re (\Xi_S
\Xi_D^*) + \frac12 |\Xi_D|^2\right\}.\label{10}\ee where $M$ is the proton
mass. It is important, that  due to vanishing of $\Xi_D$, both form factors
coincide at the threshold, as it should be (by definition), and all difference
in $\left( \left|\frac{2M G^P_E}{EG^P_M}\right|^2 -1\right)$ is due to $D$-wave
admixture.

It is interesting, that in the region $\sim 200$ MeV above threshold the
experimental ratio of $ \left|\frac{ G^\mathcal{B}_E}{G^\mathcal{B}_M}\right|,
\mathcal{B } = \Lambda, p$, shows a striking peak, which in our mechanism can
be provided by the interference of $3S$ and $2D$ resonances.  In both systems,
$\rho/\omega$ and $\phi$, theory predicts a combination  of a wide $3S$
resonance and a more narrow $2D$ resonance, see Table for $\rho/\omega$
parameters. Experimentally wide resonances in both cases are not  well
established, moreover they  are accompanied by even wider $4S$ resonances some
250 MeV higher.  Therefore to simplify calculations the BW denominator for the
$3S$ state of $\rho$ was taken as $E-E_{3S} + \frac{i\Gamma_{3S}}{2} \to i
0.15$ GeV, and  we have increased the ratio $ \frac{Re
(\Xi_S\Xi_D^*)}{|\Xi_s|^2} $ by a factor of $\sim 3$ to take into account
possible $S-D$ mixing effect on $\psi_D(0)$. The resulting ratio $ \left|\frac{
G^P_E}{G^P_M}\right|^2$ is given in Fig. 4  and agrees well  with experimental
points from \cite{4}. For the case of $\Lambda\bar\Lambda$ electroproduction
one can use the same strategy, however here the $2D$ resonance $\phi(2175)$ is
even narrower, and the ratio drops faster. We now turn to the so-called
effective form factor, defined for proton as \be |F_p|^2=\rho^2
e^{-p^2R^2_0\bar c}\frac32 \frac{E^2}{E^2+2M^2} (|\Xi_s|^2+
|\Xi_D|^2).\label{11}\ee

\begin{figure}
\caption{The ratio $\left | \frac{G^p_E}{G^p_M}\right|$ due to $3S-2D$
interference as a function of $E$. Experimental data are from \cite{4}.}
\label{misha_fig_6}
\includegraphics[width=60mm,keepaspectratio=true]{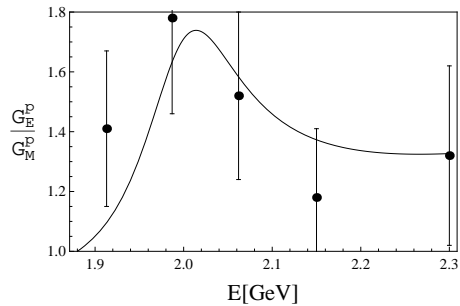}
\end{figure}

Submitting into $\Xi_S, \Xi_D$ the same combination of $3S$ and $2D$
resonances, one obtains a slowly decreasing function with $|F_p(0)|= 0.4,
\lambda=0.18$ fm. At $E=1.95$ GeV one obtains $|F_p|=0.384$ and a wide plateau
in the near-threshold region. This agrees with experimental data from \cite{4},
where the average value in the region $(1.88 \leq E\leq 1.975)$ GeV is around
0.4.  However, two lowest energy points in \cite{3,4} display a narrow ($\sim
20$ MeV) enhancement, which can be explained by the $p\bar p $  final state
interaction \cite{19}. A fast decrease of all baryon form factors and
transition to the quasiperturbative behavior \cite{5} at large $E$ requires an
explicit derivation of energy dependence of double-pair creation vertex, which
is now under study.

Summarizing, we have  applied the theory, developed before in \cite{11}, to the
cases of $\mathcal{B}\bar{ \mathcal{B}} $ electroproduction with $ \mathcal{B}=
\Lambda_c, \Lambda_b, \Lambda,p$.  The theory does not contain fitting
parameters and exploits parameters of intermediate vector mesons and baryons,
taken from calculations and experiment. The peaks above threshold are predicted
for heavy baryons $\Lambda_c, \Lambda_b$ and are in agreement with experiment
for $\Lambda_c \bar \Lambda_c$ production \cite{15}. A nontrivial angular
dependence of electroproduction is shown to occur due to interference of
$^3S_1$  and $^3D_1$ vector meson states, which also explains a nontrivial
behavior of the ratio
 $
\left|\frac{ G^{\bar {\mathcal{B}}}_E (p)}{G^\mathcal{B}_M(p)}\right| $,
observed earlier in \cite{4} for $\mathcal{B}=p$ and in \cite{16} for
$\mathcal{B}=\Lambda$. The absolute values of approximately calculated form
factors and cross sections near threshold are in general agreement with
experiment.

The author is grateful to B.O.Kerbikov for useful discussions,
  to Yu.S.Kalashnikova and  G.V.Pakhlova for discussions, useful suggestions and help,
   and to  A.M.Badalian, who provided results of calculations for heavy and light
vector mesons, given in the Table.


\begin{thebibliography}{99}

\bibitem{1}V.P.Druzhinin, S.I.Eidelman, S.I.Serednyakov and E.P.Solodov,
arXiv:1105.4975 [hep-ex].

\bibitem{2} K.K.Seth, arXiv:0712.0356 [hep-ex].

\bibitem{3}

 G.Bardin  {\it et al.} (PS 170 Collaboration), Nucl.  Phys.   {\bf B411}, 3 (1994).



\bibitem{4}

 B.~Aubert {\it et al.} ({\it BABAR} Collaboration),  Phys. Rev. {\bf D 73}, 012005 (2006).




\bibitem{5}  V.L.Chernyak and A.R.Zhitnitsky, JETP Lett {\bf 25}, 510 (1977);\\
G.P.Lepage, S.J.Brodsky, Phys. Rev. {\bf D 22}, 2157(1980);\\
 J.Bolz {\it et
al.} Z.Phys. {\bf C66}, 267 (1995);\\
 T.Hyer, Phys. Rev. {\bf D 47}, 3875
(1993).

\bibitem{6}P.Kroll, Th.Pilsner, M.Schuermann, W.Schweiger, Phys. Lett. {\bf B
316}, 546 (1993); arXiv; hep-ph 9305251; P.Kroll, Nucl. Phys. Proc. Suppl. {\bf
56A}, 33 (1997).



\bibitem{7} F.Iachello and Q.Wan, Phys. Rev. {\bf C 69}, 055204 (2004).


\bibitem{8} E. van Beveren, X.Lin, R.Coimbra et al.,  Europhys.  Let. {\bf 85},
61002 (2009);\\
M.Abud, F.Buccella and F.Tramontano, Phys. Rev. D {\bf  81}, 074018 (2010);\\
  G.Cotugno, R.Faccini, A.D.Polosa et al.,  Phys. Rev. Lett. {\bf 104},132005 (2010).


\bibitem{9} Yu.A.Simonov,Phys. Rev. {\bf D 84}, 065013 (2011).

\bibitem{10} Yu.A.Simonov and V.I.Shevchenko,  Adv. HEPh, 873051 (2009), arXiv: 0902.1405;
Yu.A.Simonov,  arXiv: 1003.3608.
\bibitem{11} Yu.A.Simonov, arXiv: 1109. 5545 [hep-ph].




\bibitem{12} Yu.A.Simonov, Phys. Atom. Nucl. {\bf 66},
338 (2003); Phys. Rev.{\bf D 65}, 116004 (2002).

\bibitem{13} A.M.Badalian, B.L.G.Bakker, I.V.Danilkin, Phys. Atom. Nucl. {\bf
72}, 638 (2009), arXiv:0805,2291 [hep-ph].

\bibitem{14} A.M.Badalian, B.L.G.Bakker, I.V.Danilkin, Phys. Atom. Nucl. {\bf
73}, 138 (2010), arXiv:0903.3643 [hep-ph].

\bibitem{14'} K.Nakamura et al (PDG), J.Phys. G: Nucl. Part. Phys. {\bf 37},
075021 (2010).

\bibitem{15}
 G.V.~Pakhlova {\it et al.} (Belle Collaboration),  Phys. Rev. Lett. {\bf 101}, 172001 (2008).







\bibitem{16} B.Aubert et al. ({\it BABAR} Collab.), Phys. Rev.  {\bf D76}, 092006 (2007).

\bibitem{17} V.A.Novikov, L.B.Okun, M.A.Shifman, A.I.Vainshtein, M.B.Voloshin
and V.I.Zakharov, Phys. Rept. {\bf 41C}, 1 (1978).

\bibitem{19} I.L.Grach, B.Kerbikov and Yu.A.Simonov,Phys. Lett. {\bf B 208},
309 (1988);

 G.Y. Chen, H.R.Dong,
J.P.Ma, Phys. Rev. {\bf D78}, 054022 (2008); O.D.Dalkarov, P.A.Khakhulin and
A.Yu. Voronin, Nucl. Phys. {\bf A833}, 104 (2010).

\end{thebibliography}
\end{document}